# Spintronic Quantum Phase Transition in a Graphene/Pb$_{0.24}$Sn$_{0.76}$Te Heterostructure with Giant Rashba Spin-Orbit Coupling

Jennifer E. DeMell*[1], Ivan Naumov[2], Gregory M. Stephen[1], Nicholas A. Blumenschein[1], Y.-J. Leo Sun[3], Adrian Fedorko[4], Jeremy T. Robinson[5], Paul M. Campbell[5], Patrick J. Taylor[6], Don Heiman[4,7], Pratibha Dev[2], Aubrey T. Hanbicki[1], Adam L. Friedman*[1]

[1] *Laboratory for Physical Sciences, 8050 Greenmead Dr., College Park, MD 20740, USA*
[2] *Howard University, Department of Physics and Astronomy, Washington, D.C. 20059*
[3] *Institute for Research in Electronics and Applied Physics, University of Maryland, College Park, MD 20740 USA*
[4] *Northeastern University, Dana Research Center, Boston, MA 02115 USA*
[5] *Electronics Science and Technology Division, Naval Research Laboratory, Washington, DC 20375 USA*
[6] *Army Research Laboratory, 2800 Powder Mill Rd., Adelphi, MD 20783, USA*
[7] *Plasma Science and Fusion Center, MIT, Cambridge, MA 02139, USA*

Corresponding authors: Jennifer E. DeMell *jdemell@lps.umd.edu*,
Adam L. Friedman *afriedman@lps.umd.edu*

**Abstract**
Mechanical stacking of two dissimilar materials often has surprising consequences for heterostructure behavior. In particular, a two-dimensional electron gas (2DEG) is formed in the heterostructure of the topological crystalline insulator Pb$_{0.24}$Sn$_{0.76}$Te and graphene due to contact of a polar with a nonpolar surface and the resulting changes in electronic structure needed to avoid polar catastrophe. We study the spintronic properties of this heterostructure with non-local spin valve devices. We observe spin-momentum locking at lower temperatures that transitions to regular spin channel transport only at ~40 K. Hanle spin precession measurements show a spin relaxation time as high as 2.18 ns. Density functional theory calculations confirm that the spin-momentum locking is due to a giant Rashba effect in the material and that the phase transition is a Lifshitz transition. The theoretically predicted Lifshitz transition is further evident in the phase transition-like behavior in the Landé g-factor and spin relaxation time.

**Introduction**

Of the many extraordinary promises of the two-dimensional (2D) materials revolution, perhaps none is as captivating as the idea that materials with widely varying properties can be arbitrarily stacked into heterostructures, regardless of lattice spacing or growth mechanism[1]. With a menu of thousands of 2D materials[2], designer heterostructures are at hand with such remarkable properties as interlayer excitons in transition metal dichalcogenides[3,4] (TMDs), "magic-angle" superconductivity in graphene and TMDs[5], and ferromagnetism from non-magnetic constituents[6]. In many of these systems, heterostructures are more than simply the sum of their parts. Rather than each layer in a stack acting as a discrete entity, unique hybridization effects result in emergent pheonomena[7,8]. One of the most versatile and useful materials is graphene: like tofu, graphene often acquires the flavor of whatever is placed in its proximity[9,10].

Perhaps the most compelling application of such heterostructures is in next-generation computing devices[11,12,13]. Primary among the many looming roadblocks of current computing paradigms are device bottlenecks in both energy expenditure and the physical limitations of the ubiquitous charge-based CMOS structures[14]. Spin-based "spintronic" devices offer considerably lower energy operation, higher speeds, and greater densities[15]. Because the spin diffusion length is proportional to mobility, a nearly defect-free graphene would seem like an ideal spintronic channel[16]. However, its lack of spin-orbit coupling leaves few options for the spin current control needed for device operation. Combining graphene with a high spin-orbit material such as a topological insulator[17,18], where the spin-orbit coupling is conveyed by proximity, could be a viable solution.

One recent example of a graphene-based heterostructure with unexpected properties is the graphene/topological crystalline insulator (TCI) system graphene/Pb$_{0.24}$Sn$_{0.76}$Te (Gr/PST). While systemic symmetries protect the properties of all topological materials, TCIs are primarily protected by mirror symmetry[19]. Since the TCI has a significantly lower conductivity than graphene, one might expect most of the current to flow through the graphene, resulting in proximitized spin-orbit coupling, similar to other graphene heterostructures[9,10,20]. However, stacking graphene onto the PST breaks the inversion symmetry and destroys the topological state. Due to excess charge, the structure must either change its stoichiometry or undergo a complete charge reconfiguration—a phenomena known as the polar catastrophe. The charge redistribution needed to avoid polar catastrophe results in the formation of a two-dimensional electron gas (2DEG) at the interface, analogous to the interface 2DEG famously discovered in LaAlO$_3$/SrTiO$_2$ and similar systems[21]. Although the topological order of the PST is destroyed in the heterostructure, electronic structure modifications as a result of electronic reconstruction due to the polar discontinuity at the interface leads to the appearance of a high-mobility 2DEG at the interface. Although this composite may not be a true topological material, it shares many properties with the parent topological PST material, such as high mobility, high spin lifetime, high spin-orbit coupling, and spin-momentum locking.

In this paper, we study the spintronic properties of the 2DEG created at the interface of graphene and PST. We fabricate and measure non-local spin valve (NLSV) devices and discover a low-temperature regime dominated by giant Rashba coupling that enables spin-momentum locking. At higher temperatures, the system transitions to a more typical NLSV device channel, attributable to a Lifshitz transition at about 40 K. Density functional theory (DFT) analysis is used to understand the spin texture and 2DEG behavior. Our DFT calculations reveal a giant Rashba spin-orbit parameter, along with the Lifshitz transition switching mechanism. We also use the Hanle effect to measure spin lifetimes as a function of temperature. We report spin lifetimes as high as 2.18 ns and spin transport persisting up to at least 500 K, properties highly desirable for spintronics applications. We also compare our measurements here with the charge transport measurements performed previously in this system. The robust spin transport and quantum phase transition in this system could be exploited for future low-power high-performance computing devices.

**Non-local spin valve measurements**

To measure the spintronic behavior of the Gr/PST system, we processed our heterostructures into standard NLSV devices (see methods below). **Figure 1(a)** shows an optical image of a measured Gr/PST spin valve and **Figure 1(b)** shows a device schematic. A charge current is applied between one outer non-magnetic reference Ti/Au contact and the adjacent inner tunnel barrier (TB)/ferromagnetic (FM) contact (injection), while monitoring the voltage across the other pair of contacts (detection). The spin injection/detection TB/FM contacts are different widths to exploit magnetic shape anisotropy. A spin-polarized charge current is injected from the FM, across the TB contact, and into the heterostructure 2DEG channel. While charge current only flows along the source-drain path, spins simultaneously diffuse in all directions. The pure spin current at the detector contact results in a spin-splitting of the chemical potential that manifests as a measurable voltage.

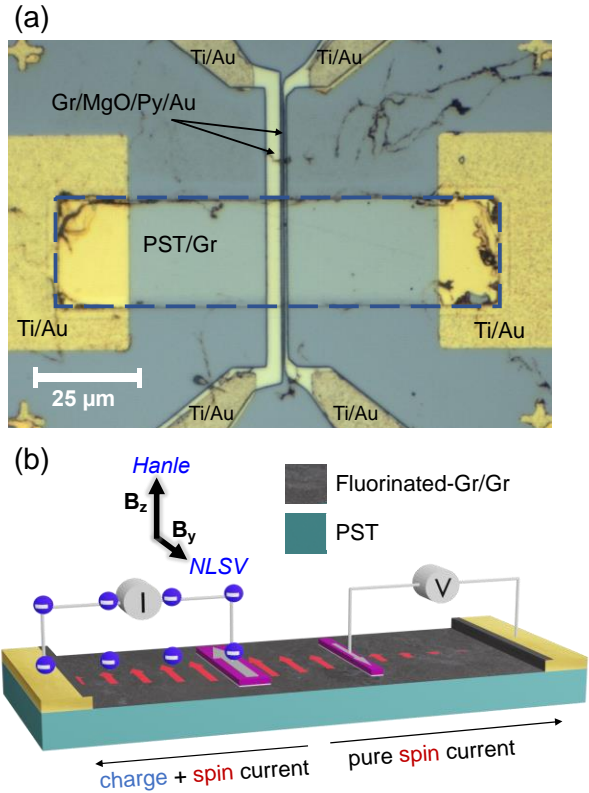

**Figure 1: Device image and operation scheme.** (a) Optical image of the Gr/PST non-local spin valve. Annotations indicate the materials used. For this device, the left and right graphene/MgO/Py/Au tunnel barrier contacts are 3 μm and 0.5 μm wide, respectively, and are separated by 1.5 μm. The channel width is 25 μm for all devices measured. (b) Device schematic of the non-local spin valve showing operation and externally applied magnetic field. As indicated, the external magnetic field is applied in-plane with the ferromagnetic contacts for non-local spin valve measurements and out-of-plane with the ferromagnetic contacts for Hanle effect measurements.

Measurement of a NLSV is essential to demonstrating spin current generation and manipulation. **Figure 2** is a summary of the NLSV measurements at various bias and temperature conditions. A sweeping magnetic field, applied along the easy axis of the FM contacts (**Figure 2(a, inset)**), switches the relative orientations of the FM magnetization at their respective coercive fields. This results in a lower (higher) measured voltage when the contacts are parallel (antiparallel). Room temperature NLSV behavior is shown in **Figure 2(a)**. The red (blue) curves are for positive (negative) magnetic field sweep directions, and the solid (dotted) lines show data acquired while applying a bias current of +10 µA (-10 µA), accounting for both spin extraction and injection. This is typical behavior for a NLSV with the difference in peak position coming from the difference in switching field of the injection/detection contacts. Switching peaks were observed up to 500 K (**Figure 2(b)**), the limit of our measurement capabilities, attesting to the robustness of the heterostructure and the device. A full set of temperature- and bias-dependent NLSV measurements are presented in the **Supplement**.

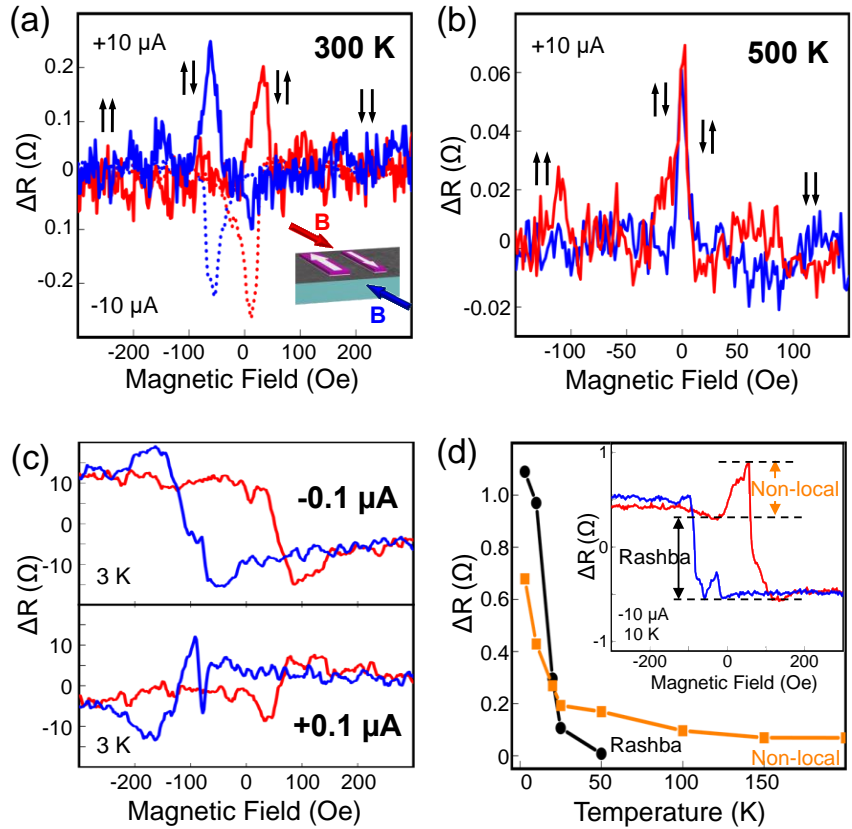

**Figure 2: Non-local spin valve device operation**. Red (blue) lines are for magnetic field sweeping negative to positive (positive to negative). A constant background resistance is subtracted for all plots. (a) Non-local resistance *vs* magnetic field at 300 K for +/- 10 µA. Inset shows a device schematic with a sweeping in-plane magnetic field. (b) Non-local resistance *vs* magnetic field at 500 K for +10 µA, demonstrating operation of the spin valve up to high temperature and the limits of our measurement equipment. (c) Non-local resistance *vs* magnetic field for +/- 0.1 µA at 3 K showing a switch in the sign of the hysteresis with bias polarity. (d) Non-local resistance of the spin injection/extraction switching peak, "Non-local," and hysteretic resistance change due to Rashba spin-orbit coupling, "Rashba," as a function of temperature. The inset shows the total non-local resistance in the spin valve *vs* magnetic field at 10 K and with a -10 µA bias and defines the relevant system resistances for analysis.

Hysteresis in the NLSV measurements of a topological or high spin-orbit coupled material is a strong indication of surface state transport. Below 50 K, the NLSV exhibits a strong hysteresis on top of the magnetization switching. This is shown in **Fig. 2(c)** with the NLSV data at 3 K. Again, the red (blue) curves are for positive (negative) magnetic field sweeps. The top (bottom) panel is a -0.1 µA (+0.1 µA) bias. Reversing the bias direction, and also the electrons' momentum, reverses the sign of the hysteresis. Similar hysteresis was also observed in both topological materials[22] and in 2DEGs[23] as a result of spin-momentum locking, though the origins of the effect for each channel type is different. For a topological material, there is a linear energy dispersion caused by band inversion where each massless, chiral Fermion has its momentum locked at a right angle to a spin-state that is thus protected by symmetry from backscattering. Our previous theory work on the PST/Gr heterostructure demonstrated that time reversal symmetry is broken here, resulting in the destruction of the topological state in the PST[21]. For a trivial 2DEG with parabolic energy dispersion, spin-momentum locking, a manifestation of spontaneous nonequilibrium spin polarization caused by an electric current, arises due to Rashba spin-orbit coupling (SOC)[24,25]. This is the same effect that allows a measure of control over spin relaxation using an electric field because the spin-splitting energy is proportional to the expectation value of electric field[26].

**Figure 2(d)** summarizes of the temperature dependence of both the non-local resistance $\Delta R_{NL}$ and the hysteretic resistance change due to Rashba spin-orbit coupling $\Delta R_{Rashba}$. The inset shows a magnetic field sweep at 10 K, with the relevant resistances defined in the annotation. $\Delta R_{Rashba}$ decreases quickly for increasing temperature until the spin-momentum locking disappears above 40 K. $\Delta R_{NL}$ also rapidly decreases with increasing temperature until equilibrating at ~150 K, where it remains up to at least 500 K (see **Supplement** for more details). Based on these findings, as well as the DFT calculations presented below, we posit a Lifshitz phase change that occurs around 40 K. This conclusion is further supported by previous measurements using only charge transport in this heterostructure[21]. In the **Supplement**, we rule out other mechanisms including a switch from ballistic-to-diffusive transport and simple thermal effects on the transport.

**Band Structure Calculations**

As the temperature changes, both the Fermi energy and the lattice parameter can also change slightly, driving the Lifshitz transition. **Figure 3** shows the calculated band structure for a Graphene/SnTe (111) interface. From the SnTe component, our system is expected to inherit topological Dirac points, which are located at the Γ- and M-points[27]. Two additional Dirac points at the Γ-point are inherited from the graphene layer. This is a consequence of zone-folding, with both the K and K' points of graphene getting folded to the Γ-point in the Brillouin zone of the heterostructure. As shown in **Figure 3(a)**, all the Dirac cones at the Γ- and M-points are gapped due to lowering of symmetry in the heterostructure. In the presence of graphene, the $C_{3v}$ point symmetry of SnTe is reduced to $C_3$-symmetry, which removes the topological protection of the surface states associated with SnTe, opening the gap. Meanwhile, the SnTe film breaks the sublattice symmetry in graphene, opening gaps in the Dirac cones associated with graphene at Γ.

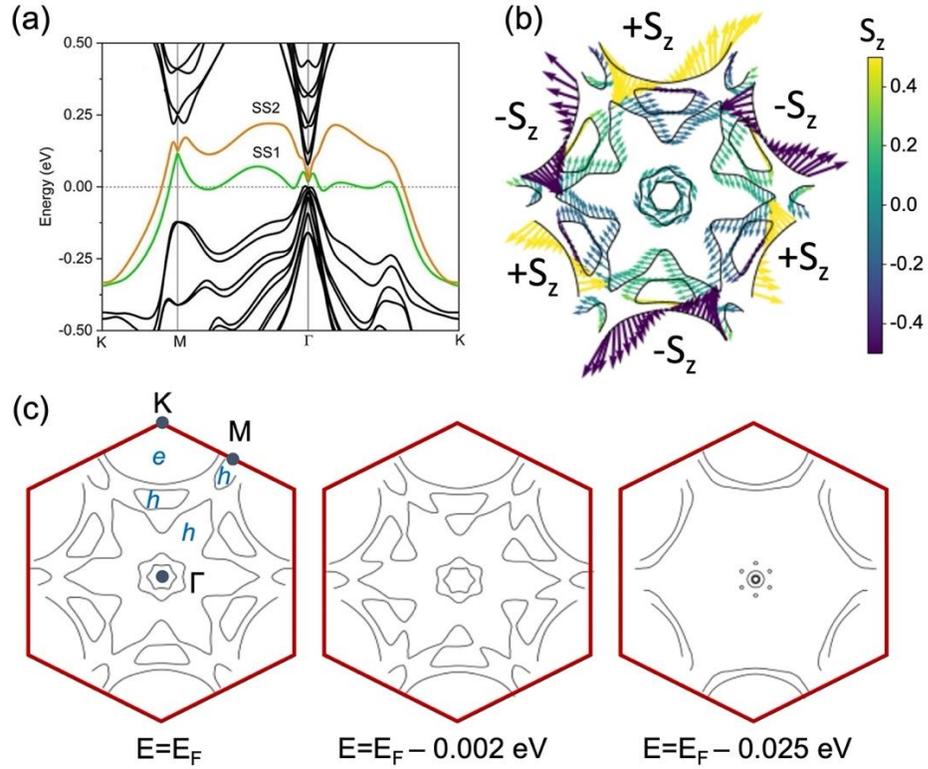

**Figure 3. (a) Calculated band structure along the high symmetry lines for the heterostructure.** Shown in orange (SS1) and green (SS2) are the two surface bands crossing the Fermi level. (b) The Fermi surfaces corresponding to the band structure presented in (a), including spin texture. (c) Constant energy contours for the energies below the Fermi level corresponding to the band structure presented in (a). Electron and hole pockets are noted by *e* and *h*, respectively.

The main feature of the band structure in **Figure 3(a)** is the presence of two surface bands crossing the Fermi level. These bands are labeled SS1 (green) and SS2 (orange). Both SS1 and SS2 are mainly localized on the upper graphene side and are characterized by both positive and negative effective masses, depending on **k**. Near the Γ point, these surface bands hybridize with the carbon $p_z$ orbitals. To estimate the strength of the Rashba effect in the heterostructure, the Rashba model was used, in which the SOC splits the parabolic band by shifting the carriers with opposite spin by the momentum $\mathbf{k}_R$. In this model, the SOC coupling parameter, $α_R$, is defined as $α_R=2E_R/k_R$, where $E_R$ is the energy difference between the crossing point and the band top or bottom at a particular $k_R$. Near the M point, along M-Γ for the SS1 band we find that $k_R ≈ 0.18$ Å$^{-1}$ and $E_R ≈ 0.13$ eV. This gives $α_R ≈ 1.5$ eV Å, which is a giant value for the Rashba parameter. For comparison, it is about half of the largest published value for the Rashba parameter in the literature for BiTeI.[28] The average Fermi velocity is also high (1.5 x 10$^5$ m/s), being nearly 30% of the Fermi velocity of Bi$_2$Se$_3$.[29]

In **Figure 3(b)**, we plot the spin-resolved Fermi surfaces (FSs) at the Fermi energy $E=E_F$. The magnitude of the in-plane component of spin is represented by the length of the arrows, and the color of the arrows gives the *z*-components of the spins. Except for the K-centered FSs, all other FSs have a small $S_z$ component and exhibit Rashba-type spin texture, with the spin being perpendicular to the **k**-vector. In contrast, the K-centered ellipses have a significant $S_z$ component. Moreover, their in-plane spin component tends to be parallel or antiparallel to **k**. Because of their large areas, the electron ellipse around K and the hole star around Γ should produce the dominant contribution to the electric current induced by an applied electric field. The **Supplement** includes calculations for the case when some of the atoms are replaced by Pb, as is the case for our exact structure.

A cross section of the band structure at various energies below the Fermi level are plotted in **Figure 3(c)**. Additional bandstructure cross sections and those above the Fermi level are found in the **Supplement**. There are multiple Fermi level crossings at different **k**-points, resulting in many FSs. The large electron ellipse around K, denoted by *e* in **Figure 3(c)**, comes from the SS2 band. There are three hole FSs, denoted by *h* in **Figure 3(c)**, including a very large star centered on Γ, all related to the SS1 band. There are also two hexagonal FSs around Γ, which are derived from graphene states.

The properties of the surface states SS1 and SS2 play an important role in the generation and registration of spin-polarized currents in the heterostructure. In our experiments, the NLSV device geometry is most sensitive to in-plane spins with Rashba-type spin-momentum locking. As the electronic pockets centered on K (associated with SS2) have high values of out-of-plane spin component $S_z$, while their in-plane spin texture is not Rashba-like in character, the current-induced spin polarization associated with SS2 is unlikely to be detected. Hence, experimentally, the role of the SS1 states is more important. The SS1 band exhibits Rashba-type spin texture, where the spin lies mostly in-plane, perpendicular to the **k**-vector. A peculiarity of the SS1 band is that it is flat in a wide region of **k**-space close to the Fermi level. As seen in **Figure 3(c)**, the SS1 FS changes rapidly for small excursions away from $E_F$. As a result, the FSs associated with this band can easily adjust their area, shape, and topology under the influence of external parameters such as pressure, composition, temperature, magnetic field, and so on. Abrupt modification of the topology of a Fermi surface is the hallmark of a Lifshitz transition. Therefore, we expect external stimuli such as temperature to strongly affect transport phenomena. For example, a large enhancement of thermoelectric efficiency in SnSe over a wide temperature range (10–300 K) was attributed to the pressure-induced Lifshitz transition[30]. Therefore, our calculations suggest that the system is on the verge of different Lifshitz transitions, which can easily occur under different stimuli, such as Pb-doping, Fermi level tuning, and lattice expansion (for example, through temperature modulation).

The number and complexities of FSs indicate that the experimentally observed disappearance of spin-momentum locking for temperatures above ~40 K can be ascribed to one or more effects. One such effect is simply the disappearance of some of the hole FSs, which as discussed have significant contribution to the spin-polarized current. Another effect is due to high anisotropy of the FSs: a measured change in resistivity, $\Delta R_{Rashba}$, should strongly depend on the orientation of the spin-induced current relative to the heterostructure. With increasing temperature, the anisotropy of the Fermi surfaces may change in such a way that the spin transfer by the drift current becomes inefficient. A third possible effect may be related to the charge redistribution between the $p_z$ orbitals of carbon and the $p$ orbitals of Sn and Te. At polar surfaces, the compensating charge density tends to persist and the carbon $p_z$ orbitals can be filled or emptied at the expense of Sn and Te $p$ orbitals[31]. Since the spin splitting of the graphene Dirac cone (whether *n*- or *p*- doped) leads to two Fermi surfaces with opposite spin directions at each momentum, the corresponding current-induced spin densities nearly cancel each other, resulting in the observed behavior.

**Spin Properties**

Another important consideration for any spintronic device is the spin polarization. In **Figure 4**, we plot the components of spin polarization derived from both the Rashba (black) and standard NLSV (red) channels. For a full discussion on the analysis of each of these components, see the **Supplement**. Comparing the Rashba polarization to the injected polarization, the Rashba component is a significant portion of the observed spin signal at low temperature before it gradually disappears.

First, we will discuss the low-temperature regime which is dominated by the Rashba effect. Although the origin of spin-momentum locking for topological materials and 2DEGs is different, the theory to analyze the hysteresis is identical[32]. For both topological materials and 2DEGs and for both ballistic and diffusive transport, the high- and low-voltage states in the hysteresis curve are described by[32–35]

$$\Delta R_{Rashba} = \frac{h^2}{e^2} \frac{P_{FM}}{2\, v_F W\, m^*} (\mathbf{p}_{Rashba} \cdot \mathbf{m_u}), \quad (1)$$

where $h$ is Planck's constant, $e$ is the elementary charge, $P_{FM}$ is the polarization of the ferromagnetic detector contact (~48% for $Ni_{80}Fe_{20}$)[36], $v_F$ is the Fermi velocity, $m^*$ is the effective mass, $W$ is the width of the channel, $\mathbf{p}_{Rashba}$ is the induced spin polarization due to the Rashba effect, and $\mathbf{m_u}$ is a unit vector along the direction of magnetization[21]. In previous studies, spontaneous, non-equilibrium spin polarization due to either the Rashba effect or topological surface state spin-momentum locking is distinguishable by having opposite signs in the vector $\mathbf{p}_{Rashba}$. However, as many others have noted[22,33,37], it is difficult to draw any conclusions about the sign of the polarization. We therefore calculate an absolute value of polarization, displayed *vs* temperature in **Figure 4.**

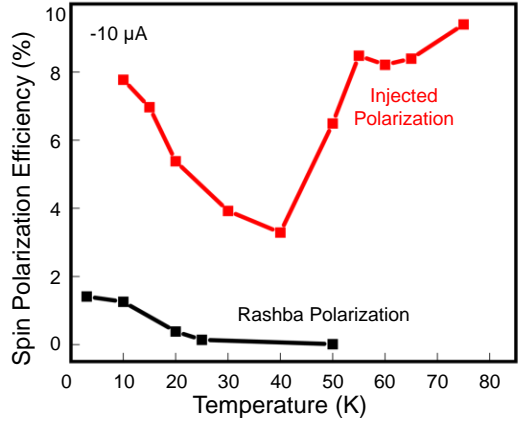

**Figure 4: Spin polarization efficiency**. Lower bound of the temperature-dependent spin polarization efficiency due to the Rashba spin-momentum locking (red) and of the non-local spin valve (black). Experimental error bars are within the size of the depicted squares at each point.

The polarization efficiency, $P_{NLSV}$, due to injection/extraction into/from a NLSV can be calculated from the injection/extraction NLSV peaks, as defined in **Figure 2(d)**. The relationship between the measured resistance and polarization[38] is $\Delta R_{NL} = \frac{P_{NLSV}^2 \lambda_s}{W\sigma} e^{-L/\lambda_s}$. Here, $W$ is the width of the spin channel, $\sigma$ is the measured temperature-dependent conductivity, and $L$ is the contact spacing. The spin diffusion length $\lambda_s = (D\tau_s)^{1/2}$, where $\tau_s$ is the spin lifetime (calculated from fitted Hanle curves below) and $D$ is the spin diffusion constant. Because our system has degeneracies (at least two spin/charge channels), we use the exact solution of the diffusion constant such that[39]

$$D = (k_B T/e)(1 + e^{-\eta_f})\ln|1 + e^{\eta_f}|, \quad (2)$$

where $\eta_f = E_g/k_B T$, $E_g$ is the system band gap obtained from DFT calculations as a lower bound on the realistic actual band gap and $k_B$ is the Boltzmann constant. For an ideal non-degenerate system, Equation (2) reduces to the Einstein relation. The spin diffusion constants calculated from a measurement of mobility in this system at 10 K for PST, graphene, and the heterostructure are 0.172 cm²/s, 3.27 cm²/s, and 17.2 cm²/s, respectively. The shape of the injected polarization in **Figure 2(d)** is due to a two-level measured spin relaxation time, discussed below. Comparing the Rashba polarization to the injected polarization, the Rashba component is a significant portion of the observed spin signal at low temperature before it gradually disappears.

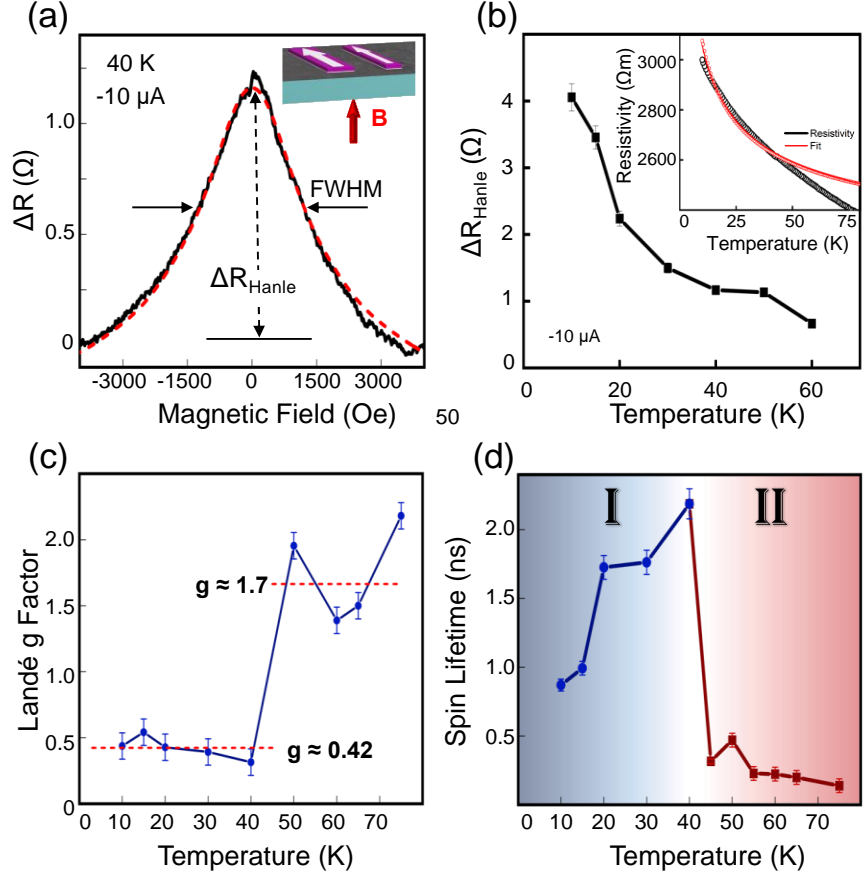

**Figure 5: Hanle Effect**. (a) Non-local resistance *vs* magnetic field of an example Hanle effect curve showing the expected pseudo-Lorentzian behavior. The red line is a fit to the solution to the spin diffusion equation using the Landé *g* factor and the spin lifetime as fitting parameters at a temperature of 40 K and bias of -10 μA. A background quadratic magnetoresistance was subtracted from the data. (b) Non-local resistance *vs* temperature for the Hanle effect until it disappears at approximately 75 K. The inset displays the four-terminal local resistivity *vs* temperature at a bias of -10 μA, fit to a model that includes both variable range hopping and thermal activation. (c) The Landé g-factor as a fitting parameter for the Hanle curves demonstrating an abrupt phase change between 40-45 K. (d) Temperature dependence of extracted spin lifetime taken at a bias of -10 μA. The two regimes corresponding to Rashba-dominated transport and spin channel-dominated transport are noted as I in blue and II in red.

Further spintronic properties of the heterostructure can be extracted from analysis of the Hanle effect, shown in **Figure 5**. Here, a magnetic field is applied out of plane with respect to the moments of the FM injector/detector contacts as is depicted in schematic in **Figure 5(a)**. This causes the spins in the channel to precess at the Larmor frequency, $\omega_L = g\mu_B B_z/\hbar$, where *g* is the Landé g-factor, $\mu_B$ is the Bohr magneton, and $B_z$ is the out-of-plane magnetic field. As the magnetic field is increased, precessional dephasing results in a higher resistance, thus a magnetic field sweep will trace out a pseudo-Lorentzian line shape proportional to solutions to the spin diffusion equation described by the steady state spin current[40]

$$S(x_1, x_2, B_z) = S_0 \int_0^\infty \frac{1}{\sqrt{4\pi Dt}} e^{-(x_2-x_1-v_d t)^2/4Dt} \cos(\omega_L t) e^{-t/\tau_s} dt, \quad (3)$$

where spin is injected into the graphene at $x_1$ and $t=0$ and is detected at $x_2$. $S_0$ is the spin injection rate and $v_d$ is the electron drift velocity (equal to 0 for pure spin currents). Fits to the Hanle data provide a measure of the spin lifetime. The Landé g-factor for graphene is known to be 2.[41] However, for PST, the Landé g-factor is not accurately known[42] and is unknown for the heterostructure 2DEG because it is difficult to separate bulk and surface g-factors in the heterostructure. Therefore, we leave the Landé g-factor as a fitting parameter.

**Figure 5(a)** is a plot of non-local resistance as a function of applied out-of-plane field for a device at a temperature of 40 K. The pseudo-Lorentzian behavior is a signature of the Hanle effect, *i.e.*, precessional dephasing of the spin current at the detection contact. The data are in black while the red dashed line shows the fit to Equation 3. A second-order polynomial background was subtracted from the raw Hanle data to remove trivial magnetoresistance. **Figure 5(b)** summarizes the change in non-local resistance *($\Delta R_{Hanle}$*, as defined in **Figure 5(a)**) as a function of temperature. Although the NLSV signal persists to at least 500 K (**Figure 2(b)**), the Hanle signal disappears above approximately 75 K. This behavior has been observed in many previously studied spin valves[43,44], with the NLSV signal far outlasting the Hanle signal, and has been attributed to thermal fluctuations and phonon scattering.

Insight into the momentum scattering behavior is learned from temperature dependent resistivity measurements. The inset of **Figure 5(b)** shows the four-terminal resistivity ($\rho$) *vs* temperature measured in this spin valve device along with a fit to a model that includes both thermal activation (TA) and Mott variable range hopping (VRH),[45] $\sigma = \sigma_{TA} + \sigma_{VRH}$. $\sigma_{TA} = A \exp[-E_a/2k_BT]$ and $\sigma_{VRH} = B \exp[-(T/T_0)^{1/3}]$. Fitting parameters are the thermal activation energy $E_a=$ 0.765 eV, temperature constant $T_0=$0.573 K, and relative conductivities $A = 0.439\ \Omega$ and $B = 221\ \Omega$. In the low-temperature regime ($T<50$ K), the VRH model dominates the temperature-dependent conduction. A stronger thermal activation or a two-state system instead of the variable range hopping was expected in this heterostructure due to the observed temperature-dependent phase change. We attribute the behavior to variable range hopping over the relevant temperature scales; however, the data does deviate from the model at higher temperatures.

Returning to a study of the spin dynamics, **Figures 5(c) and (d)** display the Hanle fit parameters $g$ and $\tau_s$ as a function of temperature. **Figure 5(c)** shows two average $g$ values; for $T <$ 40 K, $g \sim 0.42$, while for $T > 40$ K, $g \sim 1.7$. Higher temperature fits did not converge using the lower temperature $g$ value and *vice versa*. This is entirely consistent with a Lifshitz phase transition as identified above. **Figure 5(d)** shows that the extracted spin lifetime *vs* temperature also has two regimes. In regime I, the spin lifetime is higher, reaching values as large as 2.18 ns. Above 40 K, it transitions to regime II, where the spin lifetimes are markedly lower, only about 500 ps. The extracted spin lifetime *vs* temperature here was for one device, though behavior was similar for all devices tested (additional information is provided in the **Supplement**). Calculated as discussed above, the spin diffusion length $\lambda_s = 13.1$ μm for $\tau_s = 2.18$ ns. In our previous work, we used electron-electron interactions to estimate an effective spin-orbit (SO) interaction of 4.5 meV[21]. Using the values from the Hanle measurement here, we can also determine an effective SO interaction[46]. For the assumed D'yakonov-Perel' spin relaxation mechanism, $\tau_s^{DP} = h^2/(4\Delta_{SO}^2 \tau_q)$, give $\Delta_{SO} = 0.206$ meV. This value is an order of magnitude lower than our previous estimated interaction, which could be a result of interference between the Rashba and injection/extraction spin channels.

**Conclusion**

In conclusion, we demonstrated high-quality spin transport in a PST/Gr structure along with a temperature dependent Lifshitz transition. We observed spin-momentum locking in the non-local spin valve measurements due to a giant Rashba SOC, which we further calculated using DFT. Spin transport survives to at least 500 K and the devices display a long spin lifetime, long diffusion length, and high spin polarization efficiency as measured through Hanle effect measurements. The observation of a Lifshitz quantum phase transition in this heterostructure could be incorporated as the switching mechanism in a future ultra-low power computing device. Further, as results of the polar catastrophe are general for any polar/non-polar heterostructure, we predict that additional heterostructures with remarkable spintronic properties using other novel materials are imminently available and should be studied in the future.

**Acknowledgements**

The authors gratefully acknowledge support from the Office of the Secretary of Defense in an Applied Research for Advancement of S&T Priorities program (TEDs). The authors gratefully acknowledge technical support assistance at LPS from D. Crouse, P. Davis, R. Brun, G. Latini, and J. Wood. P.D. and I.N acknowledge support by the W. M. Keck Foundation and the NSF Grant number DMR-1752840. P.D. and I.N used the Bridges-2 cluster at PSC through allocation PHY180014 from the Advanced Cyberinfrastructure Coordination Ecosystem: Services & Support (ACCESS) program, which is supported by National Science Foundation Grants Nos. 2138259, 2138286, 2138307, 2137603, and 2138296, and the Maryland Advanced Research Computing Center. D.H. and A.F. thank support from NSF grant DMR-1905662 and the Air Force Office of Scientific Research award FA9550-20-1-0247. The work at the US Naval Research Laboratory was supported under base programs through the Office of Naval research.

**Competing Interests**

The authors declare no competing interests.

**Author Contributions**

A.L.F. formulated the idea and experiments. Devices were fabricated by J.E.D., A.L.F., and P.M.C. Primary electrical analysis was performed by G.M.S. Spin transport data were taken by J.E.D. with the assistance of G.M.S. and N.A.B. Data were analyzed by J.E.D. with the assistance of G.M.S. and N.A.B. and under the supervision of A.L.F. and A.T.H. DFT was performed by I.N. and P.D. J.E.D. and A.L.F. wrote the paper with assistance from G.M.S. and A.T.H, with the exception of the theoretical portions, which were written by I.N. and P.D. A.F. and D.H. performed magnetometry and magnetic analysis of the Py contacts. J.T.R. grew and transferred the graphene. P.J.T. grew the PbSnTe. Y.-J.L.S. performed optical analysis of the heterostructure stack under the supervision of A.T.H. All authors commented on the manuscript and contributed to discussions.

-------

**Methods**

**Growth Methods**

Pb$_{0.24}$Sn$_{0.76}$Te (111) (PST) topological crystalline insulator (TCI) films were grown to a thickness of 7 nm on GaAs (001) by molecular beam epitaxy using methods described previously[21]. The stoichiometry is chosen to maximize the bulk band gap while maintaining the TCI behavior, as discussed in depth elsewhere[21,47]. Graphene was grown by low pressure (5-50 mTorr) chemical vapor deposition on copper foils at 1030 °C under flowing $H_2$ and $CH_4$ gas. After growth, the Cu substrates were etched and transferred to the PbSnTe using a wet process[48].

**Device Fabrication**

To fabricate the non-local spin valve devices, we first utilized optical lithography with Shipley S1813 photoresist to pattern vias for electrical contacts on the PST followed by electron beam deposition of Ti/Au (5 nm/35nm) and lift-off in acetone. PST mesas were then defined by a second lithography step followed by ion milling in $Ar/H_2$ plasma and cleaning in an acetone/ultrasonic bath. A subsequent $O_2$ plasma descum removed any remaining photoresist residue. Two layers of graphene were then transferred on top of the PST mesas. One layer of graphene creates the graphene/PST heterostructure transport channel, while the second graphene layer is used in a tunnel barrier (TB) contact. Another lithography step using polymethyl methacrylate (PMMA) and deep-UV exposure followed by $O_2$ plasma shaped the graphene into mesas. PMMA and deep-UV lithography are used here to minimize chemical contamination of the graphene. Semiconductor NLSV devices require a TB to match the conductivities between the metallic spin injection/detection contacts and the semiconductor channel[49]. We fabricated devices with a TB consisting of fluorographene/MgO. Vias for the TB contacts are defined by optical lithography using Lift Off Resist (LOR5A) and Shipley 1813, particularly designed to minimize residue. The graphene inside the vias is fluorinated by exposure to a $XeF_2$ gas[38], which acts to both decouple the two graphene layers and selectively transform the top layer into an insulator. Immediately after fluorination, electron beam deposition is used to deposit MgO/$Ni_{80}Fe_{20}$ (Py)/Au (1.5 nm/30 nm/10 nm). An additional optical lithography/deposition step is performed to define Ti/Au (10 nm/40 nm) top contacts to ensure good electrical connection. Finally, the devices are fluorinated a second time to ensure that the top layer of graphene is completely insulating. Additional characterization details can be found the **Supplement**.

**Measurement Methods**

NLSV and Hanle effect measurements are performed in a cryogen-free variable temperature cryostat set upon a rotating platform and centered between the poles of a 1 T electromagnet. From previous work on simultaneously fabricated Hall bar devices, measured device mobilities varied from about 10,000 cm$^2$/Vs to 20,000 cm$^2$/Vs. Comprehensive charge transport data on the same heterostructures can also be found in our previous work.[21] TB contacts were evaluated using the Rowell criteria[50] and found to be of high quality, likely free of pinholes. See **Supplement** for more information. Measurements are performed with DC bias to maximize dynamic range with many measurements averaged together to eliminate noise and spurious signals.

**Theory Methods**

*Ab-initio* calculations of band structure, Fermi surfaces, and Fermi velocities for the simulated heterostructures were performed with the Vienna *Ab-initio* Simulation Package (VASP).[51,52] We adopted the PBE generalized gradient approximation (GGA) to describe the exchange-correlation potential,[53] and the Projected Augmented Wave (PAW) method[54] was used to describe the interaction between the ionic cores and electrons. Structures were relaxed until the difference in total energies between two ionic steps became smaller than $1\times10^{-4}$ eV. The van der Waals interactions between the graphene layer and the underlying thin film were accounted for by the inclusion of the many-body dispersion correction[55]. A 16x16x1 Monkhorst-Pack **k**-point grid was used for the Brillouin zone sampling of the supercell. The kinetic energy cutoff was set to 400 eV. All calculations included spin-orbit coupling (SOC). PbSnTe was modelled by replacing some of the Sn atoms with Pb atoms. However, SnTe was mainly used to approximate PbSnTe due to the expense of the calculations. Additional details can be found in the **Supplement** or in our previous work in reference 17, where we demonstrated that SnTe, PbTe, and PbSnTe all behave similarly in this context, where the dominant heterostructure behavior is from a meeting of polar and non-polar surfaces.

# SUPPLEMENTAL INFORMATION
# Spintronic Quantum Phase Transition in a Graphene/Pb$_{0.24}$Sn$_{0.76}$Te Heterostructure with Giant Rashba Spin-Orbit Coupling

Jennifer E. DeMell*[1], Ivan Naumov[2], Gregory M. Stephen[1], Nicholas A. Blumenschein[1], Y.-J. Leo Sun[3], Adrian Fedorko[4], Jeremy T. Robinson[5], Paul M. Campbell[5], Patrick J. Taylor[6], Don Heiman[4,7], Pratibha Dev[2], Aubrey T. Hanbicki[1], Adam L. Friedman*[1]

[1] *Laboratory for Physical Sciences, 8050 Greenmead Dr., College Park, MD 20740, USA*
[2] *Howard University, Department of Physics and Astronomy, Washington, D.C. 20059*
[3] *Institute for Research in Electronics and Applied Physics, University of Maryland, College Park, MD 20742 USA*
[4] *Northeastern University, Dana Research Center, Boston, MA 02115 USA*
[5] *Electronics Science and Technology Division, Naval Research Laboratory, Washington, DC 20375 USA*
[6] *Army Research Laboratory, 2800 Powder Mill Rd., Adelphi, MD 20783, USA*
[7] *Plasma Science and Fusion Center, MIT, Cambridge, MA 02139, USA*

Corresponding authors: Jennifer E. DeMell *jdemell@lps.umd.edu*,
Adam L. Friedman *afriedman@lps.umd.edu*

**Characterization of Tunnel Barriers**

In this section we characterize the tunnel barriers (TBs) for the non-local spin valve device (NLSV) in the main text. **Fig. S1(a)** shows a device schematic of the NLSV. As discussed in the main text, the outer (yellow) contacts are Ti/Au reference contacts and the inner (purple) contacts are the TB/ferromagnet stack of fluorographene/Py/Au. The two inner contacts are of different widths ("wide" and "narrow") to take advantage of shape anisotropy of the ferromagnet, both TBs are characterized independently to ensure proper tunneling. **Fig. S1(b)** shows the temperature dependence of the normalized zero-bias resistance (ZBR) with an inset of current-voltage (IV) curves across a range of temperatures (10 K up to 250 K). The non-linear IV curves and the weakly insulating behavior of the normalized ZBR suggests that there are few pinholes in the TB material, indicative of a high-quality TB[1]. **Fig. S1(c, d)** shows the IV curves for the wide and narrow TBs, respectively. The insets show the dI/dV non-linearity of the TBs, which suggests that tunneling is the dominant conduction mechanism through the TBs.

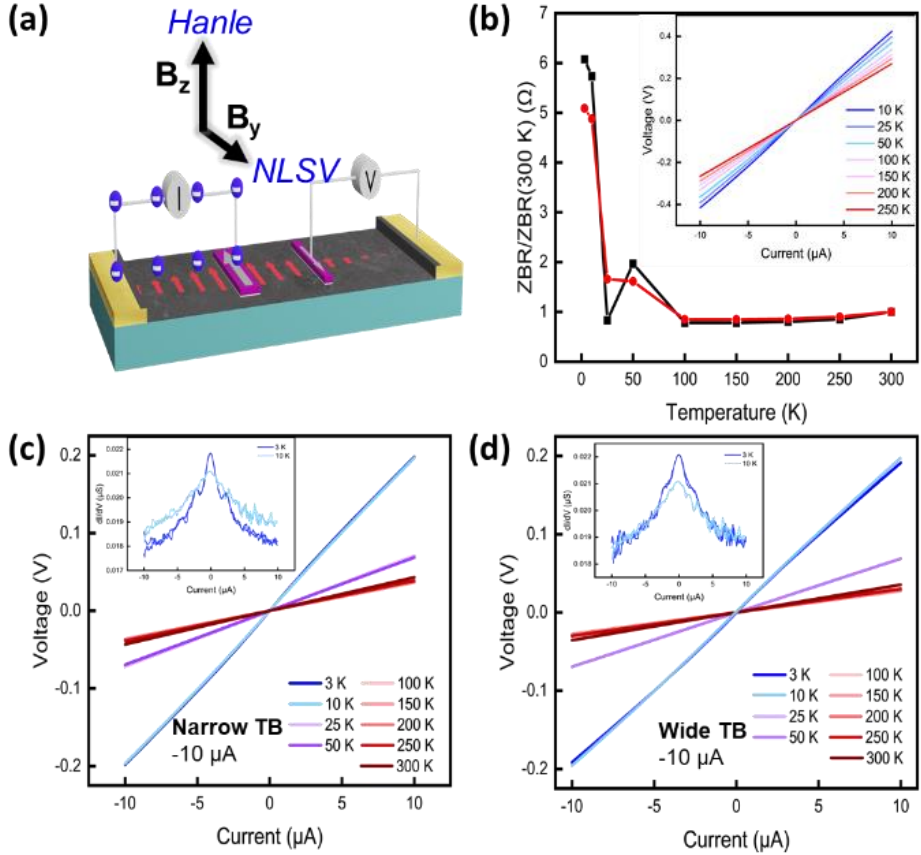

**Figure S1. Characterization of tunnel barriers.** (a) Device schematic of the NLSV. Outer (yellow) contacts are Ti/Au reference contacts and inner (purple) contacts are the wide (left) and narrow (right) ferromagnetic/TB contacts. (b) Zero-bias resistance for wide and narrow tunnel barriers. Inset shows two-terminal current *vs* bias across temperatures 10 K up to 250 K. (c) and (d) current *vs* voltage (IV) curves for wide and narrow TBs, respectively, with insets showing dI/dV curves at low temperatures.

**Additional Charge Transport Details**

The mean free path for a 2DEG with parabolic bands is given by $\ell = v_F \tau = \sqrt{2\pi n}\frac{\hbar\mu}{e}$. We calculate a mean free path of ~320 nm at 3 K. Likewise, previous measurements revealed a phase relaxation length $\sqrt{D_m \tau_\varphi}$ of ~7 nm at 3 K, with $D_m$ being the momentum diffusion constant and $\tau_\varphi$ the phase relaxation time.[17] Therefore, even at low temperature, both characteristics lengths are smaller than the device dimensions and the transport is diffusive across all measured temperatures, ruling out a switch from ballistic to diffuse transport as the cause of the observed phase change. It is possible that as temperature increases, thermal fluctuations give way to a more graphene-like channel. Because the non-local switch remains convoluted with the hysteresis at low temperatures until the hysteresis disappears at <50 K, we hypothesize that at low temperature we have a combination of spin transport from the 2DEG state and some sort of conductive bulk, perhaps due to two conductive channels crossing the Fermi energy (see the band structure calculation in **Fig. 3(a)** or **Fig. S8(a)**).

**Additional Non-local Spin Valve Analysis**

**Fig. S2** shows NLSV measurements ranging from 3 K up to 300 K with high-temperature data shown in **Fig. S3** ranging from 350 K up to 500 K. Data for **Figs. S2** and **S3** were taken with applied currents of -10 µA and +10 µA, respectively. For measurements below ~50 K, a hysteretic resistance due to Rashba spin-orbit coupling is observed in addition to the non-local resistance switching peak (see **Fig. 2** in the main text for more details). As temperatures approach the transition temperature (near 40 K), the two changes in resistance overlap and the switching is more difficult to extract. At 500 K (**Fig. S3(d)**), the highest temperature reached by our measurement equipment, the two NLSV peaks overlap, indicating that the ferromagnetic layer (Py) has softened.

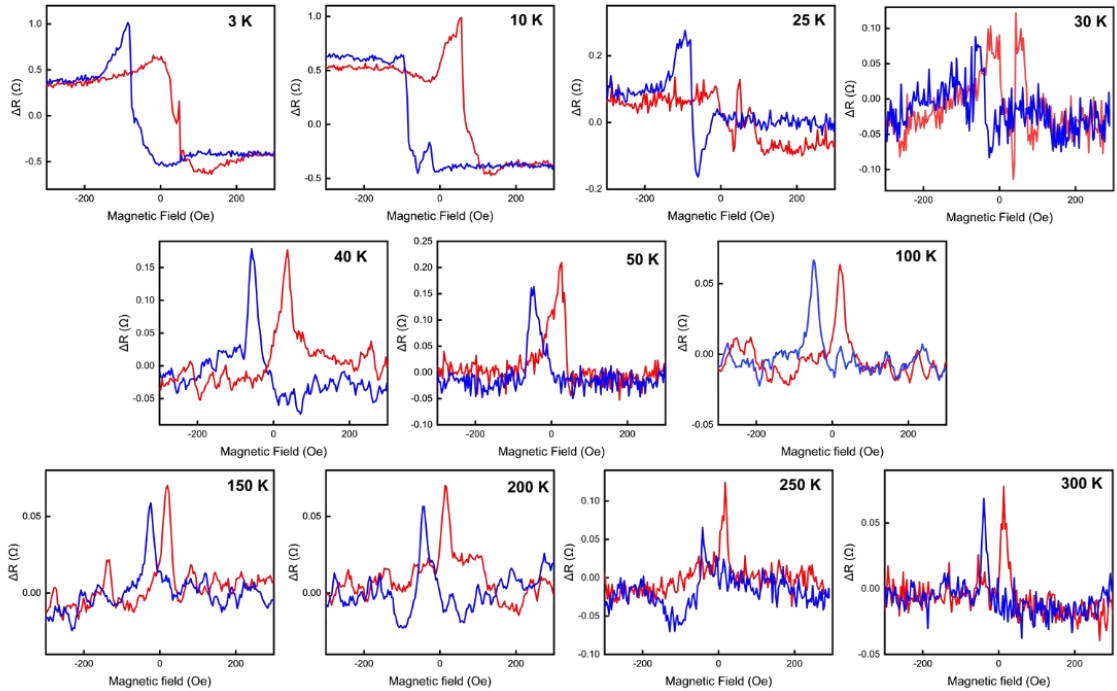

**Figure S2. Non-local spin valve measurements.** Red (blue) lines are for magnetic field sweeps from negative to positive (positive to negative). Data are shown from 3 K up to 300 K, taken at an applied current of -10 µA. A constant background resistance was subtracted for all plots.

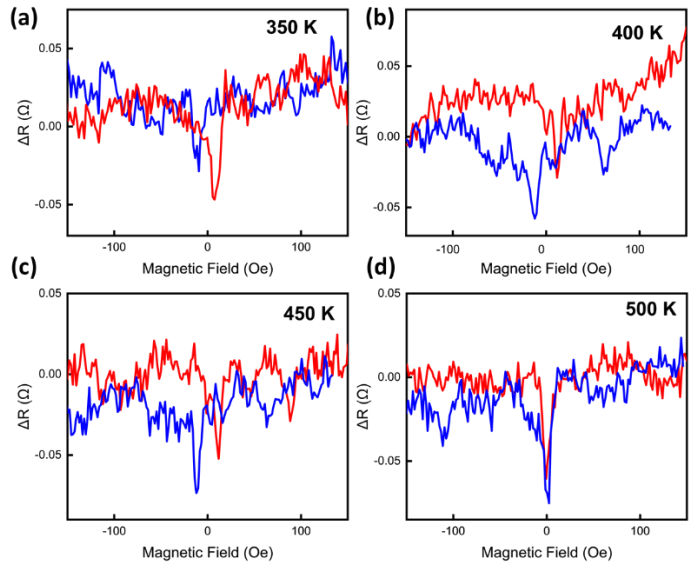

**Figure S3. High-temperature non-local spin valve**. Red (blue) lines are for magnetic field sweeps from negative to positive (positive to negative). Data are shown from 350 K up to 500 K, taken at an applied current of +10 µA. A constant background resistance was subtracted for all plots.

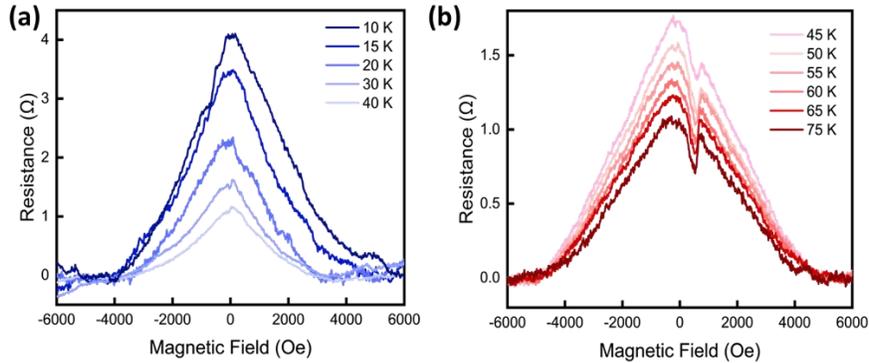

**Figure S4. Hanle data.** Hanle measurements at (a) low and (b) high temperatures taken with an applied current of -10 µA. A background subtraction was applied to remove background magnetoresistance and temperature effects.

The background-subtracted Hanle data shown in **Fig. S4 (a, b)** shows the decrease in full-width at half max of the curve as temperature increases. The feature at +600 Oe in **Fig. S4(b)** is an out-of-plane magnetic switch. This switch is symmetric when sweeping the magnetic field in either direction (±600 Oe). This feature appeared after the sample had been stored in a nitrogen dry box for approximately one year prior to additional measurements being taken and is thus likely due to some minor sample degradation with time, though the spin behavior was unaffected.

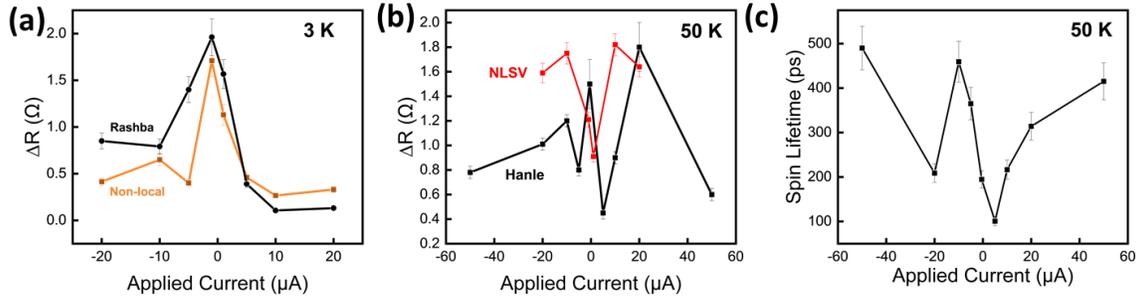

**Figure S5. Bias dependence.** Bias dependence of the non-local spin valve and Hanle measurements taken at 3 K and 50 K. (a) Change in resistance peaks for the non-local spin valve measurement configuration taken at 3 K where the orange and black points are the non-local and Rashba hysteretic changes in resistance, respectively. (b) Comparison of $\Delta R_{NL}$ at 50 K for both Hanle and NLSV device configurations. (c) Dependence of spin lifetime on bias current at 50 K.

**Fig. S5** shows that there is no clear bias dependence of the spin valve either above (50 K) or below (3 K) the transition temperature. At low temperatures, external noise and effects mask the Hanle signal making it impossible to fit to determine the spin lifetime and spin polarization efficiency values therefore data are shown only for negative applied currents (see **Fig. S5 (a-c)**.) As shown in **Fig. S6**, from the change in Rashba resistance and the spin lifetime, there is a slight increase in efficiency for low bias, similar to other studies.

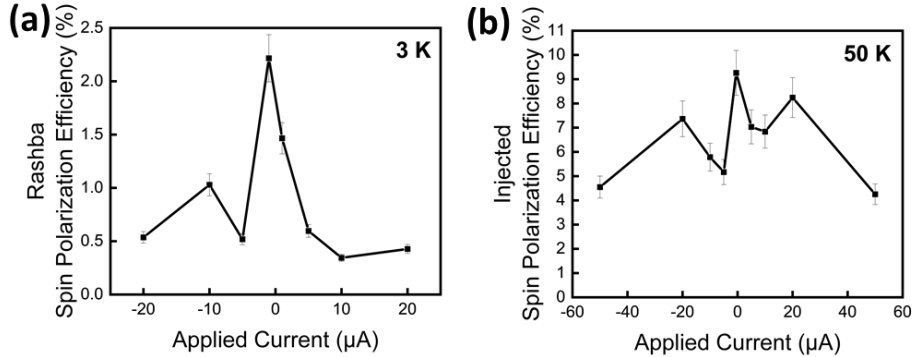

**Figure S6. Spin polarization efficiency bias dependence.** (a) Rashba hysteretic spin polarization efficiency across positive and negative bias and (b) injected spin polarization efficiency across positive and negative bias at 50 K.

Usually, a local geometry is used to study spin-momentum locking effects[2–5] due to rapid spin relaxation and owing to rapid momentum relaxation and short spin diffusion lengths in most topological materials. To measure spin diffusion through a channel, the detection contacts should be on the order of the spin diffusion length. For topological materials, it was demonstrated that spin-momentum locking could be measured in the non-local geometry because even though the spin injected into the bulk and on the top surface relaxes quickly, the induced spin polarization on the bottom surface remains detectable. The tunnel barrier contact is inherently surface sensitive; however, the creation of the 2DEG channel may greatly increase the spin diffusion length to be greater than the PST alone. The spin diffusion constants calculated from a measurement of mobility in this system at 10 K for PbSnTe, graphene, and the heterostructure are 0.172 cm$^2$/s, 3.27 cm$^2$/s, and 17.2 cm$^2$/s, respectively. Moreover, spin-momentum locking was previously measured in such a geometry owning to surface effects away from the diffusion through the bulk channel[6–8]. In any case, a non-local measurement through tunnel barrier contacts is inherently surface sensitive, although in our heterostructure we cannot disregard convolution of surface and PST bulk effects with two conductive bands crossing the Fermi energy and participating in the 2DEG transport.

**Additional Density Functional Theory Details**

To simulate the experimental heterostructures, we used a slab-supercell approach with a vacuum gap of 25 Å along the $z$ direction between the replica slabs. The basic (undoped) heterostructure was constructed by placing √3x√3 graphene supercell on a 1x1 cell of Te-terminated (111)-SnTe thin film, the bottom of which is passivated by hydrogen (**Fig. S7**). The effect of Pb doping on this basic system was modeled by replacing some Sn atoms with Pb atoms. The SnTe film was chosen to contain 35 atomic layers (17 Sn layers + 18 Te layers). Only the top four and bottom four layers of SnTe, as well as graphene and hydrogen layers, were allowed to relax fully, whereas the remaining 27 layers of SnTe were constrained to bulk positions to avoid large interlayer oscillations due to finite size effects. Calculations were performed as discussed in the main text and the methods.

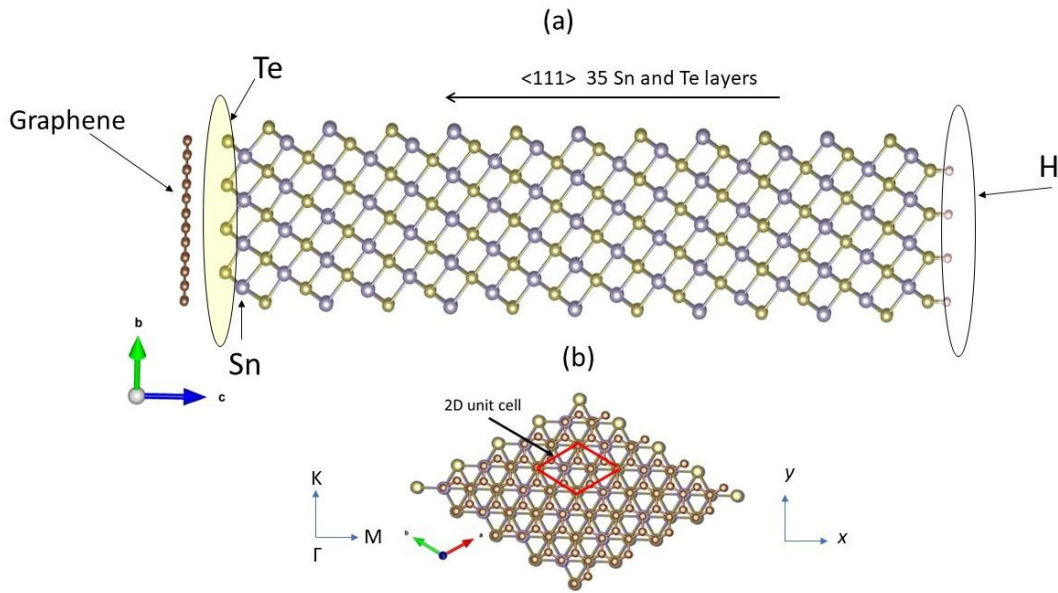

**Figure S7.** Side (a) and top (b) views of the heterostructure consisting of a √3x√3 supercell of graphene on 1x1 (111)-SnTe thin film, which is terminated in Te. The lower surface of the composite is decorated with hydrogen, passivating the Te-atoms on that surface. Also shown are the $x$ and $y$ axes in real space and the ΓM and ΓK directions in the reciprocal space.

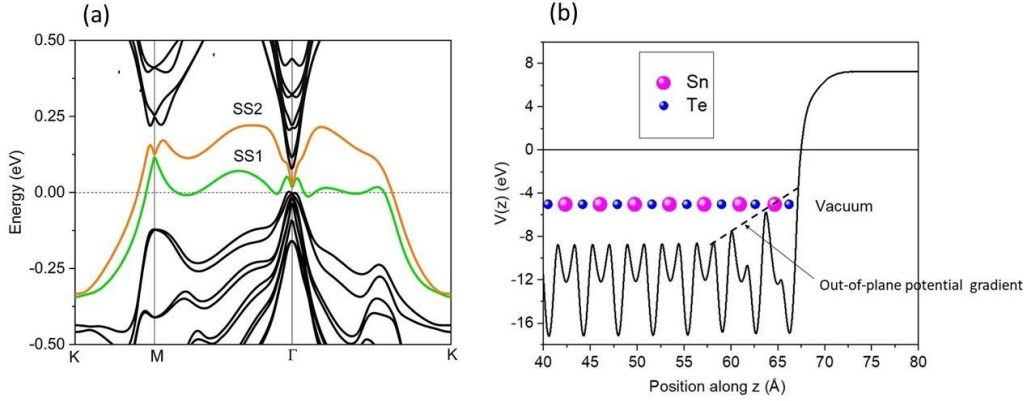

**Figure S8.** (a) Calculated band structure along the high symmetry lines for the heterostructure shown in Figure S1. Shown in orange and green (labeled as SS1 and SS2) are the two surface bands crossing the Fermi level. (b) The potential averaged over (*xy*) plane near the Te-terminated surface of SnTe, explaining the origin of the surface states SS1 and SS2 (see the corresponding text).

As discussed in the main text, the most important feature of the band structure in **Fig. S8** is the presence of two surface bands crossing the Fermi level. These bands are labeled SS1 (green) and SS2 (orange). Both SS1 and SS2 are mainly localized on the upper graphene side and are characterized by both positive and negative effective masses, depending on **k**. In the vicinity of the $\Gamma$ point, they hybridize with the carbon $p_z$ orbitals. The surface states SS1 and SS2 emerge because the (111)-surface of SnTe is polar. In the [111] direction, the sequence of atomic layers can be viewed as a sequence of charged layers +2, -2 +2, -2 …, if we assign to the Sn and Te ions their formal charges equal to +2 and -2. In our system, the polarity is canceled by introducing one more Te layer on the bottom, which violates the stoichiometry. This new layer leads to the formation of the surface states SS1 and SS2, which provide the needed, positive compensating charge density $\sigma_{ext}$ (1 hole per unit cell area) on both sides[17]. The system is metallic because together SS1 and SS2 surface bands must be approximately half-filled. The specific features of SS1 and SS2 can be understood looking at the potential averaged over (*xy*) plane near the Te-terminated surface of SnTe (see **Fig. S8(b)**). The potential bends near the surface becoming more positive or less binding. This explains why the surface states SS1 and SS2, split off from the uppermost bulk valence bands. On the other hand, the potential gradient near the surface is responsible for the giant spin-splitting between SS1 and SS2 states.

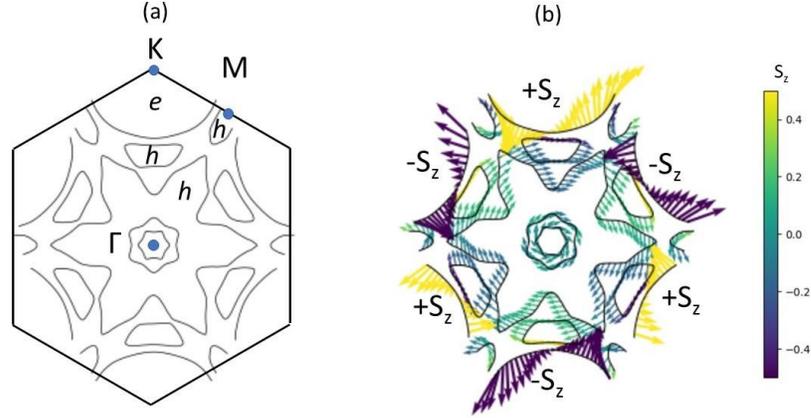

**Figure S9.** The Fermi surfaces corresponding to the band structure presented in Figure S8, without (a) and with (b) spin texture.

The band structure in **Fig. S8(a)** shows several bands crossing the Fermi level at different **k**-points. This results in many Fermi surfaces (FSs) plotted in **Fig. S9(a)**. The large electron ellipse around K comes from the SS2 band. There are three hole FSs, including a very large star centered on Γ. All of them are related to the SS1 band. There are also two hexagonal FSs around Γ, which are derived from graphene states. On the right in **Fig. S9(b)**, we plot the spin-resolved FSs. The magnitude of the in-plane component of the spins is indicated by the length of the arrows, whereas the color of the arrows is used to give the *z*-components of the spins. With the exception of the K-centered FSs, all other FSs have a small $S_z$ component and exhibit Rashba-type spin texture, with the spin being perpendicular to the **k** vector. In contrast, the K-centered ellipses have a significant $S_z$ component. Moreover, their in-plane spin component tends to be parallel or antiparallel to **k**. It is reasonable to assume that because of their large areas, the electron ellipse around K and hole star around Γ should produce the dominant contribution to the electric current induced by an applied electric field.

When all Sn atoms are replaced by Pb atoms in the heterostructure, we observe large changes in shape and topology of the FSs. Comparing **Figs. S9** and **S10**, we can conclude that upon such an atomic substitution, the two small hole FSs (at M and between Γ and K) disappear, whereas the star-like hole pocket transforms into a corrugated hole circle. In addition, since the states around Γ move up in energy, more Dirac-like cones centered on Γ cross the Fermi level. At the same time, the K-centered electron pockets change their shape only slightly and mostly retain the original spin texture. The above comparison suggests that the original system (without Pb) is on the verge of different Lifshitz transitions, which can easily occur under different stimuli, such as Pb doping, Fermi level tuning, and lattice expansion.

To illustrate this point, in **Figs. S11** and **S12** we present constant energy contours for the energies just below and above the Fermi level, respectively for the basic (undoped) heterostructure. One can notice multiple topological changes in the isosurfaces, especially below the Fermi level, where the first Lifshitz transition occurs already at ~0.001 eV below $E_F$, and represents spillover of holes from two pockets into one.

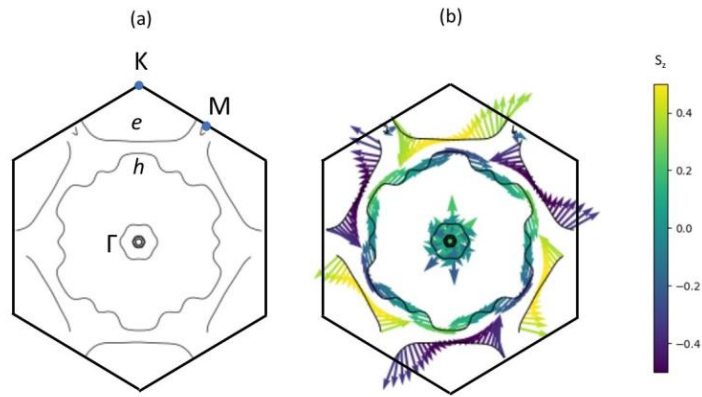

**Figure S10.** Same as in Figure S9, but for the heterostructure where the SnTe film is replaced by PbTe film.

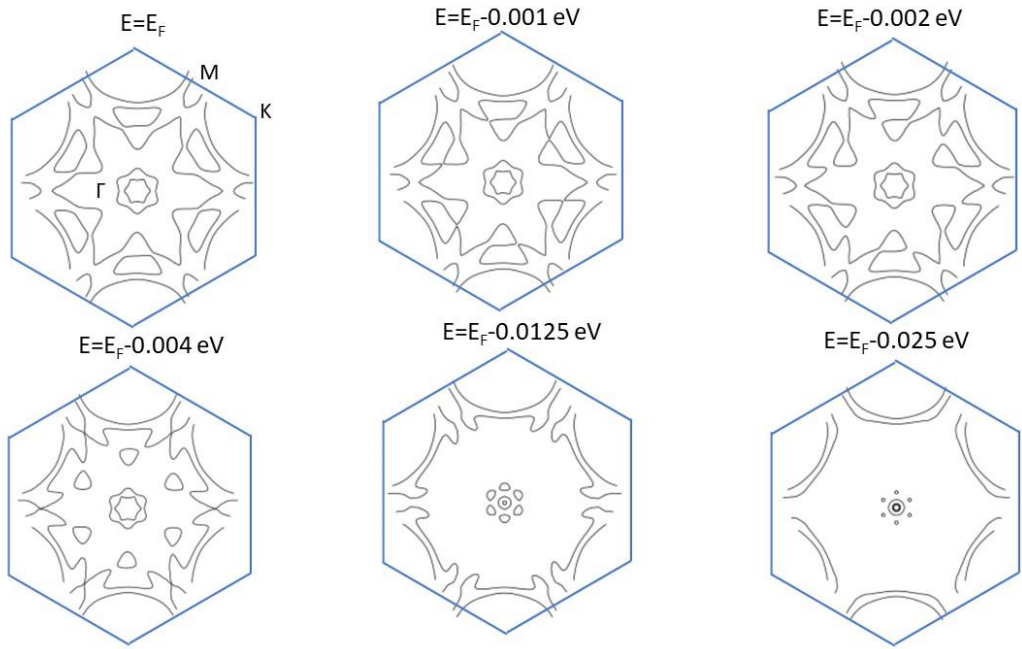

**Figure S11.** Constant energy contours for the energies below the Fermi level corresponding to the band structure presented in Figure S8.

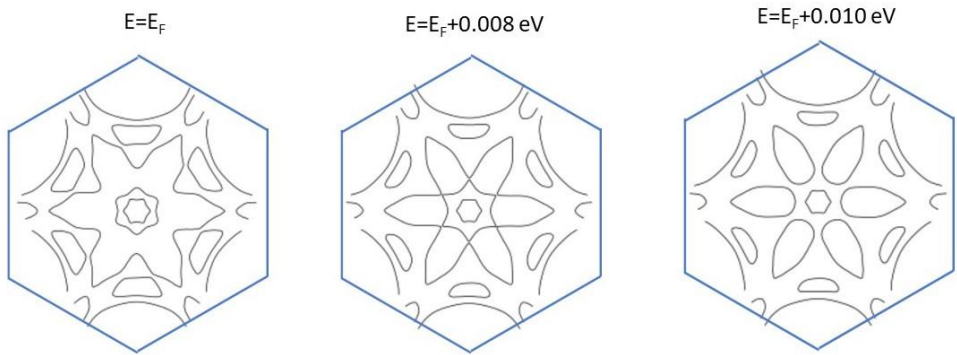

**Figure S12.** Same as in Figure S11, but for the energies above the Fermi level.